\documentclass[%
aps,
prl,
superscriptaddress,
reprint,
twocolumn,
amsmath,
amssymb,
floatfix,
]{revtex4} 

\usepackage{graphicx}
\usepackage{dcolumn}
\usepackage{bm}
\usepackage{amssymb,amsmath}
\usepackage{times} 
\usepackage[upint]{newtxmath} 
\usepackage[T1]{fontenc}

\begin{document}

\preprint{APS/123-QED}

\title{Intelligence of small groups}

\author{Giovanni Francesco Massari}
\author{Garland Culbreth}
\author{Roberto Failla}
\affiliation{Center for Nonlinear Science, University of North Texas, P.O. Box 
311427, Denton, Texas 76203-1427, USA}

\author{Mauro Bologna}
\affiliation{Departamento de Ingenier\'ia El\'ectrica-Electr\'onica, 
Universidad de Tarapac\'a, Arica, Chile}

\author{Bruce J. West}
\affiliation{Army Research Office, CDCC US Army Research Laboratory, Research 
Triangle Park, NC 27708, USA}

\author{Paolo Grigolini}
\affiliation{Center for Nonlinear Science, University of North Texas, P.O. Box 
311427, Denton, Texas 76203-1427, USA}

\date{\today}

\begin{abstract}
Dunbar hypothesized that $150$ is the maximal number of people with whom one
can maintain stable social relationships. We explain this effect as being a
consequence of a process of self-organization between $N$ units leading
their social system to the edge of phase transition, usually termed
criticality. Criticality generates events, with an inter-event time interval
distribution characterized by an inverse power law (IPL) index $\mu _{S}<2$.
These events break ergodicity and we refer to them as crucial events. The
group makes decisions and the time persistence of each decision is given by
another IPL distribution with IPL index $\mu _{R}$, which is larger than  
$\mu _{S}$ if $N\neq 150$. We prove that when the number of interacting
individuals is equal to $150$, these two different IPL indexes become
identical, with the effect of generating the Kardar Parisi Zhang (KPZ)
scaling $\delta =1/3$. We argue this to be an enhanced form of intelligence,
which generates efficient information transmission within the group. We
prove the information transmission efficiency is maximal when $N=150$, the
Dunbar number.
\end{abstract}

\maketitle


The social brain hypothesis of Dunbar \cite{1992,socialbrainhypothesis} has
created substantial interest due to its connecting of cognition with
sociology. The number of people with whom a single individual can establish
stable social relations is the "magic" \cite{bahcall} number 150 and is
related by Dunbar to the connectivity of neurons within the brain.
Physicists have historically introduced such "magic numbers" to highlight
important patterns in complex data sets for which no underlying theory had
yet been established, but for which the need was evident.

G. West, in a recent popularization of his two decades of collaborative
research on allometry relations \cite{geoffrey}, illustrated the attempts at
explaining the Dunbar hypothesis including the conjecture that the number
150 may reflect the desire of individuals within a group to maximize their
assets while realizing the maximal filling of social space. The still more
recent book of Bahcall \cite{bahcall} relates the Dunbar effect to the
occurrence of phase transitions in complex phenomena. Bahcall conjectures
that, as far as complexity management issues are concerned, the long-term
survivability of a group, company, or organization, depends on its adopting
the form of organization that supports innovative ideas favoring societal
progress. The recent work of Mahmoodi \emph{et al} \cite{sotc} proposes a
theory by which a complex network can realize self-organization, and
provides a theoretical foundation supporting the speculations of both G.
West and Bahcall just cited. According to the self-organized temporal
criticality (SOTC) model \cite{sotc} the process of self-organization is
determined by the actions of single individuals to receive the largest
individual payoffs, and they accomplish this by simultaneously reaching
maximal agreement between their opinion and the opinions of their their
nearest neighbors. The direct calculation of the transmission of information
from one SOTC system to another was found to be non- monotonic with
increasing network size. The maximum information transfer occurred when the
number of units $N$ of the two self-organizing systems is in the range [$%
100,200]$.

The SOTC proposed by Mahmoodi et al. \cite{sotc} shifts the focus from the
complexity of amplitude avalanches \cite{bak} to the complexity of time
intervals between consecutive events \cite{book2019}. The spontaneous
transition of complex dynamics to criticality generates crucial events that
are responsible for the sensitivity of the network dynamics to the
environment. In addition crucial events entail the maximization of
information transport from one to another SOTC system. The time interval $%
\tau $ between consecutive crucial events is given by waiting-time
probability density functions (PDFs) sharing in the intermediate asymptotic
regime \cite{intermediate} the same inverse power law (IPL) structure as:

\begin{equation}
\psi (\tau )=(\mu -1)\frac{T^{\mu -1}}{(T+\tau )^{\mu -1}},
\label{manneville}
\end{equation}%
with $1<\mu <3$.
Note that the IPL structure holds only in the intermediate asymptotic regime
being exponentially truncated in the long-time regime. This truncation
limits the efficiency of information transmission and it is important to
explain the magic number $N=150$. In fact, decreasing $N$ has the effect of
not only increasing the intensity of temporal fluctuations, which favors 
system's sensitivity, but also that of decreasing the extension of the complex 
intermediate asymptotics, which yields the opposite effect. 

In the last few years the conjecture has been made that biological systems
function best when their dynamics are close to criticality \cite{bialek}.
Thus, it is reasonable to implement the associations made among complexity,
criticality and collective behavior to address the issue of cognition using
the concept of a collective mind \cite{couzin}. Long-range correlations are
amplified at the onset of a phase transition and are often studied by means
of the Ising model \cite{cardy}. On the other hand, the Ising model at
criticality generates intermittent \cite{contoyannis} and crucial events 
\cite{shuster}, which according to Paradisi and co-workers \cite{paradisi}
is a manifestation of consciousness.


The criticality hypothesis is widely shared in neuroscience \cite%
{plenz1,plenz2,chialvo,cavagna}, but its quantitative implementation
requires further advances in the renormalization group formalism. The
systems requiring study are characterized by having only a small number of
units \cite{bialekveryrecent} $N$, thereby entailing a theory of criticality
of small groups necessary for its understanding \cite%
{chialvochallenging,cavagnachallenging}. Do these results conflict with the
concept of universality, or are they compatible with the discovery of a new
universality class? It is convenient to quote the work of Takeuchi and Sano 
\cite{liquidcrystal} on the growth of liquid crystals with the discovery of
the scaling index $\delta =1/3$, namely the Kardar-Parisi-Zhang (KPZ)
scaling \cite{kpz}. Takeuchi and Sano find the same turbulent effects as
those created by Ising criticality and the same scaling as that generated by
the random growth of surfaces through the model of ballistic deposition \cite%
{failla}, thereby generating the question of a possible connection between
the Ising and KPZ universality classes.

In this Letter we address the issue of consciousness and complexity by
exploring the connection between criticality and the Dunbar hypothesis. We
do this by relating the consciousness/complexity issue to the use of scaling
theory in the search for the origin of anomalous diffusion series $\xi (t)$ and 
using a mobile window to tranform the fluctuations characterized by $\xi (t)$ 
into many diffusional trajectories $x(t)$. The purpose of their procedure was 
to establish that the departure of $\xi (t)$ from a completely random function, 
could be detected through the departure of the scaling of $x(t)$ from ordinary 
diffusion by means of a scaling index\ being different from $\delta =0.5$.

We make the statistical analysis of time series generated by two models of
criticality- induced intelligence, with a method recently proposed to detect
crucial events by Culbreth et al. \cite{garland}. This method shares the
same purpose as that of an earlier paper \cite{scafetta}, based on
converting the time series data into a diffusion process and is called
Diffusion Entropy Analysis (DEA) to determine the scaling of a diffusive
process. When criticality-induced intelligence becomes active the
constructed process is expected to depart from ordinary diffusion signified
by having a scaling index different from $\delta =0.5$. The modified DEA
(MDEA) illustrated in \cite{garland} overcomes the limits of the original
DEA technique \cite{scafetta} that were pointed out by Scafetta and West 
\cite{scafettawest}. As noticed in the latter reference the original version
of DEA cannot assess if the deviation from $\delta =0.5$ is due to the
action of crucial events, or to the infinite memory contained in Fractional
Brownian Motion (FBM) \cite{fbm}. The present analysis, using the new, MDEA,
filters out the scaling behavior of infinite stationary memory, when it
exists, and the remaining departure of the scaling index from $\delta =0.5$
is thereby solely a consequence of crucial events.

The MDEA applied to the signal $\xi (t)$ generated by the
criticality-induced intelligence implements the original DEA in conjunction
with the Method of Stripes (MoS). In the MoS the $\xi $-axis is divided into
many bins of equal size $s$ and an event, either crucial or not, is detected
when $\xi (t)$ moves from a given stripe to an adjacent stripe. A random
walker (RW) step is triggered by such an event and the RW makes a step of
constant length forward each time an event occurs, thereby generating a
diffusional trajectory $x(t)$ affording information on the opinion
persistence. To detect this information we apply the MDEA method also to 
$x(t)$.

We have selected two models generating criticality-induced intelligence. The
first is the Decision Making Model (DMM) \cite{book} where $N$ individuals
have to make a choice between two conflicting decisions. They do that under
the influence of their nearest neighbors. This model falls into the Ising
universality class, thereby making it possible for us to compare our results
to the predictions of Chialvo \cite{chialvo}. The second is the model of
swarm intelligence proposed by Vicsek and co-workers \cite{vikseck}. We
evaluate the intelligence of the DMM system using both the signal $\xi (t)$
and its time derivative $\eta \equiv d\xi /dt$. In both cases we select the  
values of the control parameters yielding criticality when the number of units 
is much larger than the Dunbar number $N= 150$ and we monitor their dynamics  
for $N$ moving from these large values to values smaller than $N=150$.

\begin{figure*}[t!]
  \begin{center}
    \includegraphics[width=1.00\linewidth]{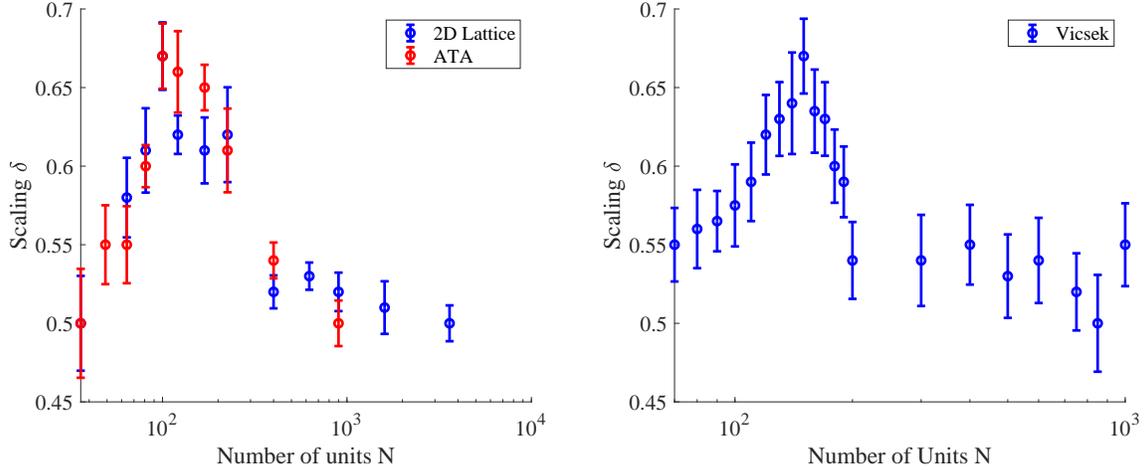}
  \end{center}
  \caption{Scaling detection of the Dunbar number is obtained by calculating
  the non-montonic dependence of the scaling index $\protect\delta $ on a
  network of size $N$. On the left two calculations are depicted using a DMM 
  \cite{book}; the red points with an all-to-all interaction, the blue
  points with a nearest neighbor interaction on a two-dimensional lattice. On
  the right the same calculation is carried out using the model of swarm
  intelligence proposed by Vicsek and co-workers \cite{vikseck}.}
  \label{Fig1}
\end{figure*} 

Fig. \ref{Fig1} illustrates the results of that analysis. The qualitative
agreement between DMM, on the left panel, and swarm intelligence, on the
right panel, is remarkable, and in the supplementary material we show that
the model of ballistic deposition \cite{failla} shares the same qualitative
agreement. The scaling $\delta $ is very close to the value $\delta =0.67$
when $N$ is in the vicinity of the magic number $150$ and falls quickly to $%
\delta =0.5$ on the left, for $N<150$ and more slowly to the same value on
the right, for $N>150$. One is tempted to interpret this result as a sign
that intelligence emerges only at $N=150$, but that would be premature.

\begin{figure}[h]
  \begin{center}
    \includegraphics[width=1.00\linewidth]{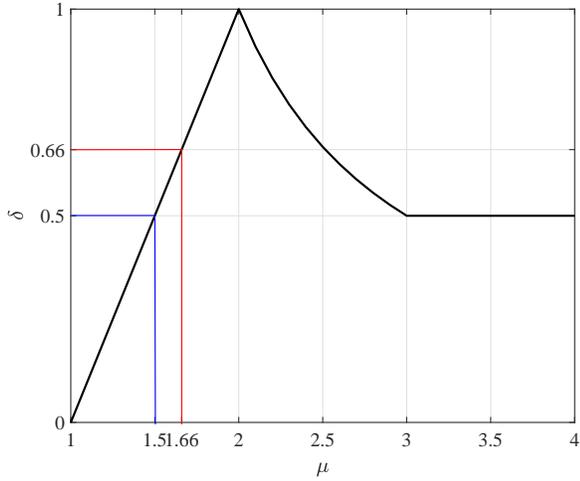}
  \end{center}
  \caption{Sketch of the connection between the scaling $\protect\delta $ and
  the crucial event power index $\protect\mu _{S}$ detected by using DEA with
  stripes, that is, MDEA.}
  \label{Fig2}
\end{figure}

It is well known that a diffusion trajectory generated by totally random
fluctuations yields a rare recursion to the origin and the time distance
between consecutive origin crossings is described by Eq. (\ref{manneville})
with $\mu =1.5$ \cite{redner}. We notice however, that the distance between
consecutive origin re-crossings affords information about the system keeping
the same opinion, while MDEA applied to $\xi $ detects the IPL of crucial
events. Therefore, it is convenient to use the symbol $\mu _{R}$ to denote
the complexity of opinion persistence and the symbol $\mu _{S}$ to denote
the temporal complexity of crucial events. Due to always making a step
forward we have \cite{giacomo}: $\delta =\mu _{S}-1$ for $1<\mu _{S}<2$; $%
\delta =1/(\mu _{S}-1)$ for $2<\mu _{S}<3$ and $\delta =0.5$ for $\mu _{S}=3$%
. See Fig. \ref{Fig2}.

According to this rule $\delta =0.5$ is generated by both $\mu _{S}=1.5$ for 
$\mu _{S}<2$ and by $\mu _{S}>3$. The second condition is equivalent to
ordinary Poisson processes. In conclusion, $\delta =0.5$ may be determined
by both crucial events with $\mu _{S}=1.5$ and non-crucial events, namely,
by $\mu _{S}>3$.

In the case when the time series $\xi (t)$, with positive and negative
fluctuations, is generated by crucial events only, we obtain the Continuous
Time Random Walk (CTRW) \cite{weiss}, along with the scaling, which can be
properly evaluated using DEA without stripes: $2\delta =\mu _{S}-1.$ In this
case $\mu _{R}=1+\delta $ (see supplement). Under the strict condition that
both crucial events and opinion persistence are renewal processes we obtain 
\begin{equation}
\mu _{R}=2-\frac{\mu _{S}-1}{2},  \label{fromStoR}
\end{equation}%
a relation originally proposed by Failla \emph{et al} \cite{failla} to study
the random growth of surfaces.

\begin{figure}[h]
  \begin{center}
    \includegraphics[width=1.00\linewidth]{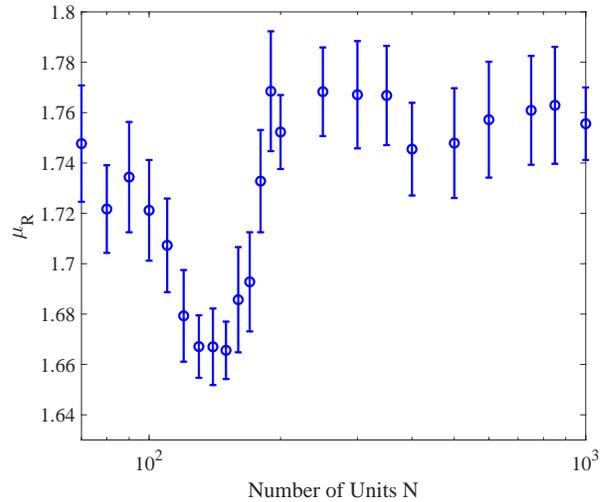}
  \end{center}
  \caption{$\protect\mu_R$ as a function of $N$ in the case of swarm
  intelligence \cite{vikseck}.}
  \label{Fig3}
\end{figure}

Thus, in cases of $N<150$ or $N>150$, where $\mu _{S}=1.5$, we expect that $%
\mu _{R}=1.75$. To evaluate $\mu _{R}$ we study the diffusional variable 
\begin{equation}
x(t)=\int_{0}^{t}dt^{\prime }\xi (t^{\prime })+x(0).
\end{equation}%
The diffusional variable $x(t)$ spends an extended time in the region $x>0$,
corresponding to the system selecting the "yes" state, and an extended time
in the region where $x(t)<0$, corresponding to the system selecting the "no"
state. This is the opinion persistence effect, previously mentioned. For
this reason evaluating the IPL index $\mu _{R}$ is a challenging
computational problem that we overcome by applying the MDEA to $x(t)$. In
this latter case the scaling $\delta $ evaluated by MDEA affords $\mu _{R}$
as $\mu _{R}=1+\delta $. This scaling $\delta $ is different from the
scaling obtained by observing $\xi $, but the value of $\mu_{R}$ should be
identical to the observation of the regression to the origin of $x(t)$.
These results for $\mu _{R}$ are shown in Fig. \ref{Fig3} wherein the
dependence of $\mu _{R}$ on $N$ is depicted.

\begin{figure*}[t]
  \begin{center}
    \includegraphics[width=1.00\linewidth]{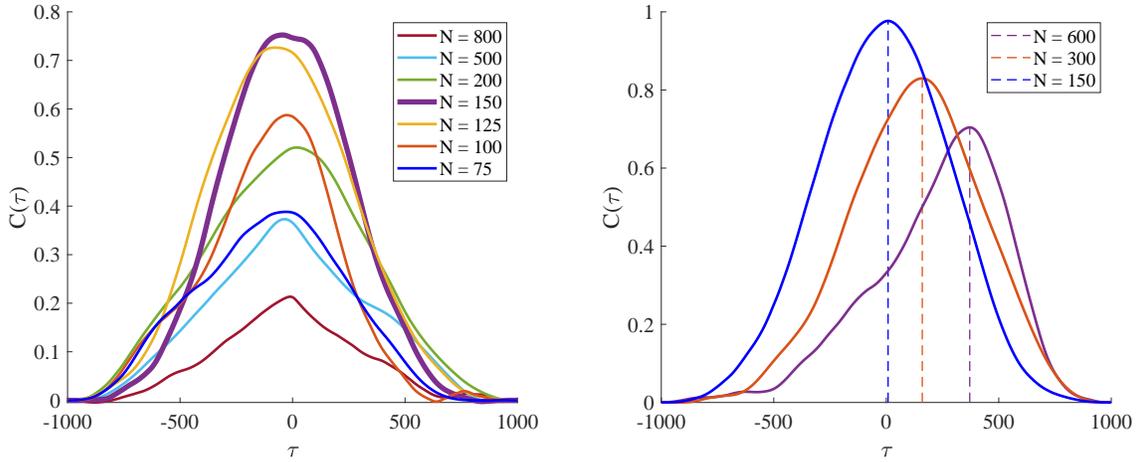}
  \end{center}
  \caption{The cross-correlation function of two interacting flocks, $A$ and $%
  B $, at criticality. Left panel: 5 \% of units of $A$ determine their flying
  direction by selecting the average flying direction of their $6$ nearest
  neighbors and of one bird of B. 5 \% of units of $B$ determine their flying
  direction by selecting the average flying direction of their $6$ nearest
  neighbors and of one bird of $A$. Right panel: the system $A$ is not
  influenced by the system $B$. }
  \label{Fig4}
\end{figure*}

In Fig.\ref{Fig3} for values of $N$ either significantly smaller, or
significantly larger, than $150$, $\mu _{R}$ is close to $1.75$, whereas at $%
N=150$, $\mu _{R}$ is close to the value $\mu _{R}=1.67$. This figure
supports the conjecture that changing $N$ affects $\mu _{S}$, making it move
from $\mu _{S}=1.5$ to the value $\mu _{S}=\mu _{KPZ}\equiv 5/3\approx 1.67$%
. This value of 5/3 is obtained from Eq. (\ref{fromStoR}) under the
condition that $\mu _{R}=\mu _{S}$, which makes the SOTC scaling of $2\delta
=\mu _{S}-1$ yield the KPZ scaling $\delta =1/3$. In other words, the Dunbar
effect makes the IPL index of the crucial events identical to the IPL index
of opinion persistence. The results of Fig. \ref{Fig3} establishes that
approaching the Dunbar number makes the crucial events located at the
position illustrated by the blue vertical line of Fig. \ref{Fig2} move to
the position of the red vertical line of of the same figure. There are still
non-crucial events in the region of Fig. \ref{Fig2} with $\mu _{S}>3$, but
the MDEA filters these out and perceives only the genuinely crucial events
moving from $\mu _{S}=1.5$ to $\mu _{S}=\mu _{KPZ}=5/3$.

The left panel of Fig. \ref{Fig1} contains the global fields of DMM with the
All-To-All (ATA) condition is realized using the equation \cite%
{stochasticdmm} 
\begin{equation}
\eta \equiv \dot{\xi}=g \sinh K(\xi +f(t)) - g \xi \cosh K(\xi +f(t)),
\label{fundamentalequation}
\end{equation}%
where $1/g$ defines the time scale of the single units, when they work in
isolation. The random noise $f(t)$ has intensity proportional to $1/\sqrt{N}$, 
thereby affording information on the size of the complex system.

The important question of why the processes of self-organization make the
systems evolve toward the KPZ condition must be addressed. Why does
self-organization favor processes with no distinction between $\mu _{R}$ and 
$\mu _{S}$? We have remarked earlier that the time interval between
consecutive crossings of the origin can be interpreted as the opinion
persistence time. Thus, we conjecture that tuning $\mu _{S}$ to $\mu _{R}$
facilitates information transport. We prove that this is the advantage of
the KPZ condition: It improves the information transfer efficiency.


Let us consider two complex systems (flocks of birds) $A$ and $B$ at
criticality. They are identical, having the same number of units $N$, and
they interact with one another. For a time $L$ the global field $\xi (t)$ of
the system $A$ and the field $\zeta (t)$ of the system $B$ are correlated. We 
evaluate the cross correlation 
\begin{equation}
C(\tau )=\frac{\int_{0}^{L-\tau }dt(\xi (t)-\bar{\xi})(\zeta (t+\tau )-%
\overline{\zeta })}{\sqrt{\int_{0}^{L}dt(\xi (t)-\bar{\xi}%
)^{2}\int_{0}^{L}dt(\zeta (t)-\overline{\zeta })^{2}}},
\end{equation}%
where $\bar{\xi}$ and $\overline{\zeta }$ denote the time averages of the
field $\xi (t)$ and $\zeta (t)$, respectively.

This cross-correlation experiment is done in two ways. In the first case a
small percentage, $5\%$ of units of $A$, randomly chosen, make their choice
on the basis of the choices made by their nearest neighbors and one randomly
chosen unit of the system $B$. The system $B$ is influenced by the system $A$
through the same interaction process. As a result of this back-to-back
interaction the cross-correlation time is expected to be symmetric around $%
\tau =0$. In the second case, we expect that the cross-correlation function
shifts to the right as a consequence of the fact that the information about $%
A$ transmitted by $5\%$ of $B$ units perceiving the motion of $A$ does not
have the immediate effect of making all the other units adopt the motion of $%
A$. The authors of \cite{cavagna} made the conjecture that the
transmission of information from the lookout birds to all the other birds of
the system occurs through a diffusion process. Lukovic and co-workers \cite%
{noinformationwave} argued that the change of direction of the flock
requires a sufficiently large number of origin re-crossings, namely, they
assigned to the visible crucial events an important role in information
transmission. We direct the readers' attention to the right panel of Fig.
\ref{Fig4}. It shows that this time delay between driven and driving
networks is extremely small when $N=150$, an evident sign that the Dunbar
effect facilitates the transport of information from the lookout birds (the
5\%) to all the other birds of the flock.

\emph{Conclusion}.--- A system at criticality generates visible crucial events,
that is the changes of opinion with the IPL index $\mu _{R}$. These events have 
a IPL index that may be different  from $\mu_S$, the IPL index  of hidden 
crucial events \cite{invisible}, see Eq. (\ref{fromStoR}). Specifically $
\mu_R \geq \mu_S$, see Fig. 5 of \emph{supplementary material}. The MDEA makes 
it possible to evaluate the scaling generated by the hidden crucial events. 
Fig. \ref{Fig3} shows that at $N=150$ the visible events adopt the same 
temporal complexity as the invisible crucial events. Another technique we use 
to make the crucial event visible is the adoption of the time derivative of $
\xi $, yielding the results illustrated by the left panel of Fig. \ref{Fig1}. 
This alternative approach leads to the same conclusion that the Dunbar 
hypothesis is realized by $\mu _{R}=\mu _{S}$, a property yielding the KPZ 
universality class, and corresponding to maximal sensitivity of the complex 
system to its environment.

We believe that the results of this Letter contribute to progress on the
scaling universality \cite{universality1,universality2,prosen}. The Dunbar
effect may activate a special form of out of equilibrium dynamics, since the
KPZ scaling $\delta = 1/3$ \cite{prosen}, interpreted as determined by the
action of crucial events, yields the KPZ power index $\mu_{KPZ} = 5/3$. 
$\mu_{KPZ}$ is identical to Kolmogorov power index, which according to 
\cite{universality1} favors in the pre-scaling regime \cite{universality2} the
energy distribution across the scales.

\emph{Acknowledgments}: The authors thank the U.S. Army Research Office for
supporting this research work through grant W911NF1901.

\bibliography{IntelligenceofSmallGroups} 

\end{document}


\author{Giovanni Francesco Massari}
\author{Garland Culbreth}
\author{Roberto Failla}
\affiliation{Center for Nonlinear Science, University of North Texas, P.O. Box 
311427, Denton, Texas 76203-1427, USA }

\author{Mauro Bologna}
\affiliation{Departamento de Ingenier\'ia El\'ectrica-Electr\'onica, 
Universidad de Tarapac\'a, Arica, Chile }

\author{Bruce J. West}
\affiliation{Information Science Directorate, US Army Research Office, Research 
Triangle Park, NC 27708, USA}

\author{Paolo Grigolini}
\affiliation{Center for Nonlinear Science, University of North Texas, P.O. Box 
311427, Denton, Texas 76203-1427, USA }

\date{\today}

\begin{abstract}

\end{abstract}

\title{\textit{Supplementary Material for}\\ "Intelligence of small groups"}

\maketitle

\appendix
\section{Trajectories}

The MDEA applied to $\xi(t)$ determines $\mu_S$ and applied to $x(t)$ evaluates 
$\mu_R$. Theoretically, the connection between $\mu_R$ and $\mu_S$ is 
established by
\begin{equation} \label{roberto}
\mu_R = 2 - \frac{\mu_S -1 }{2},
\end{equation}
a relation proposed in Ref. \cite{failla} to account for the results obtained 
from the ballistic deposition model.

The proof of Eq. (\ref{roberto}) is based on the scaling property of diffusion 
processes
\begin{equation}
p(x,t) = \frac{1}{t^{\delta}} F(\frac{x}{t^{\delta}}).
\end{equation}
Assuming that all the trajectories of a Gibbs system are located on the origin 
$x = 0$ at $t=0$, we have

\begin{equation} \label{summation}
p(0,t) = \sum_{n=1}^{\infty} \psi^{(R)}_n(t) = \frac{1}{t^{\delta}} F(0), 
\end{equation}
where $\psi^{(R)}_n(t)$ is the probability that a trajectory starting from the 
origin at time $t=0$ returns to the origin $n$ times, with the last return 
occurring at time $t$. It is well known that if the recrossing of the origin is 
a renewal process, adopting the Laplace transform yields:

\begin{equation}
\hat \psi^{(R)}_n(u)= (\hat \psi^{(R)}(u))^n .
\end{equation}
Thus, by Laplace transforming Eq. (\ref{summation}) and making the proper 
geometric summation we obtain

\begin{equation} \label{intermediate}
\hat p(0,u) = \frac{\hat \psi^{(R)}(u)}{1 - \hat \psi^{(R)}(u)} \propto 
\mathcal{L} (1/t^{\delta}). 
\end{equation}
For the Laplace transforming operation we use the notation

\begin{equation}
\hat f(u) = \mathcal{L} f(t).
\end{equation}
The Laplace transform of $\psi^{(R)}(t) \propto 1/t^{\mu_R}$ with $\mu_R < 2$ 
is $\hat \psi^{(R)}(u) = 1 - \Gamma(2-\mu_R) u^{\mu_R- 1}$, thereby making the 
intermediate term of Eq. (\ref{intermediate}) proportional to $1/u^{\mu_R-1}$. 
The Laplace transform of $1/t^{\delta}$, due to the Tauberian theorem, is 
proportional to $1/u^{1-\delta}$. Thus, Eq. (\ref{intermediate}) yields $\mu_R 
= 2 - \delta$, which, using the Continuous Time  Random Walk (CTRW) scaling 
\cite{failla}

\begin{equation} \label{traditional}
\delta = \frac{\mu_S -1}{2},
\end{equation}
leads to Eq. (\ref{roberto}). 

\section{Subordination of a Merely Diffusion Process}

The authors of Ref. \cite{failla}, in addition to the earlier trajectory 
argument, described the process of random surface growth by means of the 
subordination theory. The subordination of a merely diffusion process is 
described by the following time convoluted Fokker-Planck equation

\begin{equation} \label{onlydiffusion}
\frac{\partial}{\partial t} p(x,t) = D \int_{0}^t dt' \Phi(t-t') 
\frac{\partial^2}{\partial x^2} p(x,t'),
\end{equation}
where the memory kernel $\Phi(t)$ is defined through its Laplace transform

\begin{equation}
\hat \Phi(u) = \frac{u \hat \psi_{S}(u)}{1 - \hat \psi_{S}(u)}.
\end{equation}
The waiting time distribution density $\psi_S(\tau)$ in the intermediate 
asymptotics is proportional to $1/\tau^{\mu_S}$, with $\mu_S < 2$, thereby 
playing the role of embedding crucial events into the diffusion process. Here 
we show that in this case

\begin{equation} \label{casea}
\mu_R = 1 + \delta, 
\end{equation}
where $\delta$ is given by Eq. (\ref{traditional}). 

We do our calculation with the assumption that 
\begin{equation}
\hat \psi_S(u) = \frac{1}{1 + \left(\frac{u}{\lambda}\right)^{\alpha}},
\end{equation}
where 

\begin{equation}
\alpha \equiv \mu_S - 1. 
\end{equation}
This waiting time distribution density corresponds to a survival probability 
with the structure of a Mittag-Leffler function and with $\alpha < 1$ it has 
the asymptotic  behavior $1/\tau^{\mu_S}$, with $\mu_S < 2$ that we hypothesize 
in this paper. 

We work with the Mittag-Leffler function defined as a power series in the time 
representation

\begin{equation}\label{nd1}
E_{\alpha}(-t^\alpha)=\sum_{n=0}^{\infty} \frac{(-1)^{n}t^{n\alpha}}{\Gamma(n
\alpha+1)}. 
\end{equation}
The Laplace transform of its derivative is

\begin{equation}\label{nd2}
{\cal L}\left[ - \frac{d}{dt} E_{\alpha}(-t^\alpha) \right]= \frac{1}
{u^{\alpha}+1},
\end{equation}
where for sake of simplicity we set time scale parameters to unity. The time 
derivative of the Mittag-Leffler function is positive for $0<\alpha<1$ (or $1<
\mu_S<2$). Returning to Eq. (\ref{onlydiffusion}) with

\begin{equation}\label{nd5}
p(x,0) = \delta(x).
\end{equation}
Taking the Laplace-Fourier transform of (\ref{onlydiffusion}) we obtain

\begin{equation}\label{nd5}
\hat{p}(k,s)= \frac{1}{D \hat{\Phi}(u)k^2+u} .
\end{equation}
Note that the adoption of the Mittag-Lefler function yields
\begin{equation}\label{nd3}
\hat{\Phi}(u)=\frac{u\hat{\psi_S}(u)}{1-\hat{\psi_S}(u)} =u^{1-\alpha},
\end{equation}

\begin{equation}\label{nd6}
\hat{p}(x,u)= \frac{e^{-\frac{\sqrt{u} \left| x\right| }{\sqrt{D\hat{\Phi}
(u)}}}}{2 \sqrt{Du\hat{\Phi}(u)} }.
\end{equation}

The Laplace transform of first time distribution $f(t)$ is given by  
\cite{weiss}

\begin{equation}\label{ftp1}
\hat{f}(u)= \frac{ \hat{p}(x,u)}{ \hat{p}(0,u)}=\exp\left[-\frac{\sqrt{u} 
\left| x\right| }{\sqrt{D\hat{\Phi}(u)}}\right]=\exp\left[-\frac{u^{\alpha/2} 
\left| x\right| }{\sqrt{D}}\right].
\end{equation}
We obtain this result by supplementing the Weiss prescription \cite{weiss} with 
Eq. (\ref{nd3}) and Eq. (\ref{nd6}). The anti-Laplace transform of Eq. 
(\ref{ftp1}) yields the following asynptotic behavior,

\begin{equation}\label{ftp2}
f(t)\approx \frac{ c \mid x\mid }{t^{\frac{\alpha}{2}+1} },
\end{equation}
which proves Eq. (\ref{casea}).

\section{Subordination to a Fluctuation-Dissipation Process}

In this case the time convoluted Fokker-Planck equation to discuss is:

\begin{equation} \label{withfriction}
\frac{\partial}{\partial t} p(x,t) =  \int_{0}^t dt' \Phi(t-t') \left[ \gamma 
\frac{\partial }{\partial x} x + D\frac{\partial^2}{\partial x^2}  
\right]p(x,t').
\end{equation}
We show that this physical condition yields

\begin{equation} \label{impressive}
\mu_R = \mu_S. 
\end{equation}
In this case we again assume the initial condition of Eq. (\ref{nd5}). Taking 
the Laplace transform of (\ref{withfriction}) we obtain

\begin{equation}\label{caseb2}
u\hat{p}(x,u)-\delta(x)= \hat{\Phi}(u)\left[\gamma\frac{\partial}{\partial x }x
+ D\frac{\partial^2}{\partial x^2}\right]\hat{p}(x,u),
\end{equation}
which for $x\neq 0$ gives

\begin{equation}\label{caseb20}
u\hat{p}(x,u)= \hat{\Phi}(u)\left[\gamma\frac{\partial}{\partial x }x+ D
\frac{\partial^2}{\partial x^2}\right]\hat{p}(x,u). 
\end{equation}
Solving the above equation we have

\begin{equation}\label{sol}
\hat{p}(x,u)= c_1 e^{-\frac{\gamma x^2}{2 D}} H_{-\frac{u}{\gamma \hat{\Phi}
(u)}}\left(\frac{\sqrt{\gamma } x}{\sqrt{2} \sqrt{D}}\right)+c_2 e^{-
\frac{\gamma x^2}{2D}} \, _1F_1\left(\frac{u}{2 \gamma \hat{\Phi}(u)};\frac{1}
{2};\frac{x^2 \gamma }{2 D}\right),
\end{equation}
where $H_{\nu}(z)$ is the Hermite function and $ _1F_1\left(a;b;z\right)$ is 
the confluent hypergeometric function. We impose the following conditions

\begin{eqnarray}\label{caseb31}
&&\hat{p}(x,u)\to 0,\,\, \textrm{for}\,\, x\to\pm\infty,
\\\label{caseb32}
&&\hat{p}(0^-,u)=\hat{p}(0^+,u),
\\\label{caseb33}
&& \frac{\partial \hat{p}(x,u)}{\partial x}\mid_{x=0^+}-\frac{\partial \hat{p}
(x,u)}{\partial x}\mid_{x=0^-}=-\frac{1}{ D\hat{\Phi}(u) }.
\end{eqnarray}
Condition (\ref{caseb33}) is obtained integrating Eq. (\ref{caseb2}) around 
$x=0$. We have

\begin{equation}\label{caseb4}
\hat{p}(x,u)=\frac{\sqrt{\gamma } e^{-\frac{\gamma x^2}{2 D}} 2^{\frac{u}
{\gamma  \hat{\Phi} (u)}-\frac{1}{2}} \Gamma \left[\frac{1}{2} \left(\frac{u}
{\gamma \hat{\Phi} (u)}+2\right)\right] H_{-\frac{u}{\gamma \hat{\Phi} (u)}}
\left(\frac{\sqrt{\gamma } \left| x\right| }{\sqrt{2} \sqrt{D}}\right)}
{\sqrt{\pi } \sqrt{d} u}.
\end{equation}
Note that $ \hat{p}(x,u)$ given by Eq. (\ref{caseb4}) is normalized, i.e. $
\int_{-\infty}^{\infty} \hat{p}(x,u)dx=1/u$. Using for $\hat{\Phi}(u) $ 
expression (\ref{nd3}), for the first time distribution we have

\begin{equation}\label{ftp1b}
\hat{f}(u)= \frac{ \hat{p}(x,u)}{ \hat{p}(0,u)}= \frac{2^{\frac{u^a}{\gamma }} 
e^{-\frac{\gamma x^2}{2 D}} \Gamma \left(\frac{u^a+\gamma }{2 \gamma }\right) 
H_{-\frac{u^a}{\gamma }}\left(\frac{\sqrt{\gamma } x}{\sqrt{2} \sqrt{D}}
\right)}{\sqrt{\pi }}.
\end{equation}
Taking the limit for $u\to 0$ and for $x$ small enough we have

\begin{equation}\label{ftp1b}
\hat{f}(u)\approx 1-
\frac{u^{\alpha} \left[2 H'\left(0,\frac{\sqrt{\gamma } x}{\sqrt{2} \sqrt{d}}\right)+\gamma_E \right]}{2 \gamma },
\end{equation}
where $H'\left(0,z\right)$ is the derivative of the Hermite function with 
respect to the index $\nu$ at $\nu=0$, and $\gamma_E$ is the Euler gamma. In 
the time representation

\begin{equation}\label{ftp1c}
f(t)\approx \frac{c(x)}{t^{1+\alpha}}=\frac{c(x)}{t^{\mu}},
\end{equation}
which proves Eq. (\ref{impressive}).

\section{Scaling evaluation through MDEA}

\begin{figure}[ht]
\centering
\includegraphics[width=0.5\linewidth]{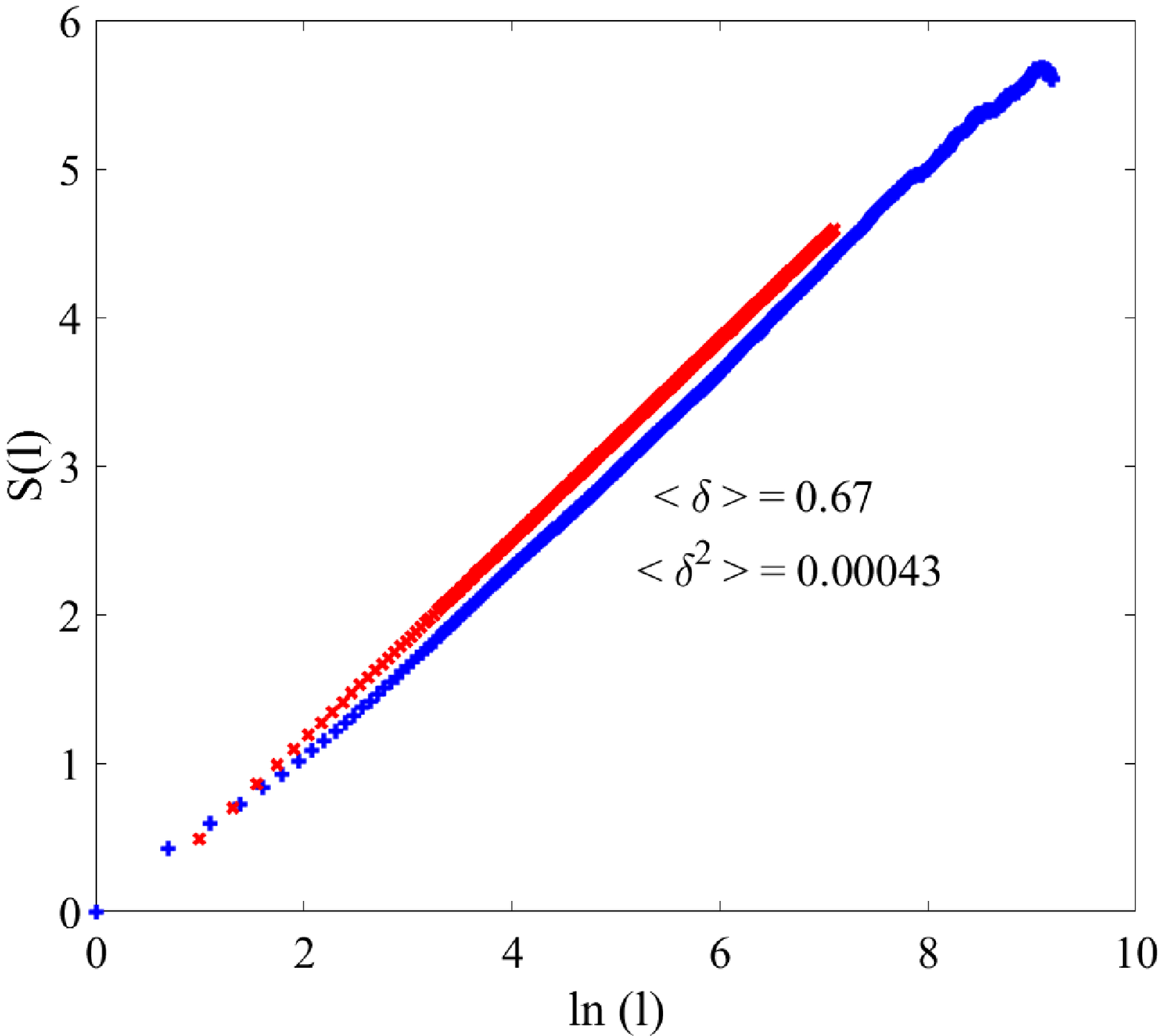}
\caption{MDEA applied to the DMM model in the ATA condition, Eq. (4) of the 
text. } \label{ATA}
\end{figure}
Fig. \ref{ATA} shows how the scaling $\delta$ is  defined. This figure refers 
to the specific case of ATA illustrated in Fig. 1 of the main text. It is the 
slope of the red straight line. Its extension affords information about the 
temporal length of the intermediate asymptotic regime. The reason why Fig. 3  
of the main text has large errors is due to the fact  in that case the 
extension of the intermediate asymptotics is not very prolonged.

\section{MDEA Applied to CTRW}

\begin{figure}[ht]
\centering
\includegraphics[width=1.00\linewidth]{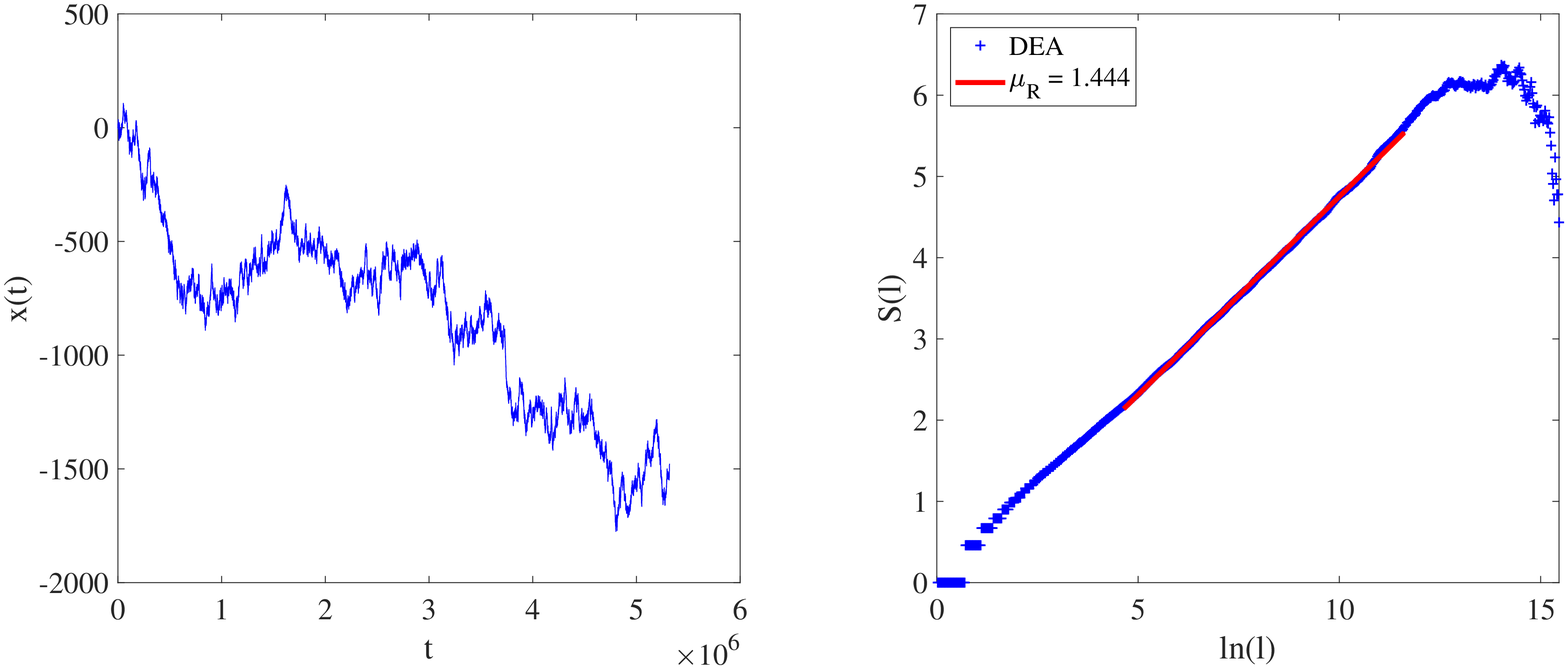}
\caption{Left panel: the diffusion trajectory $x(t)$ generated by CTRW with $
\mu_{S} = 1.888$; right panel: $S(l)$ generated by MDEA applied to the 
diffusional trajectory of the left panel.}
\label{fig1}
\end{figure}

This paper shows that MDEA applied to the diffusion trajectories generated by 
the fluctuations $\xi$, driven by crucial events with $\nu_S < 2$, yields a 
reliable evaluation of the parameter $\mu_R$. We apply this technique to the 
case of fluctuations $\xi(t)$ generated by CTRW. In this case the laminar 
regions between two consecutive crucial events are empty, thereby making the 
procedure more accurate. 

In Fig. \ref{fig1} we show the method in action on a CTRW signal with $\mu_S = 
1.8888$. We see that the theoretical prediction $\mu_R = 1.444$ is recovered 
with good accuracy. 

\begin{figure}[ht]
\centering
\includegraphics[width=0.5\linewidth]{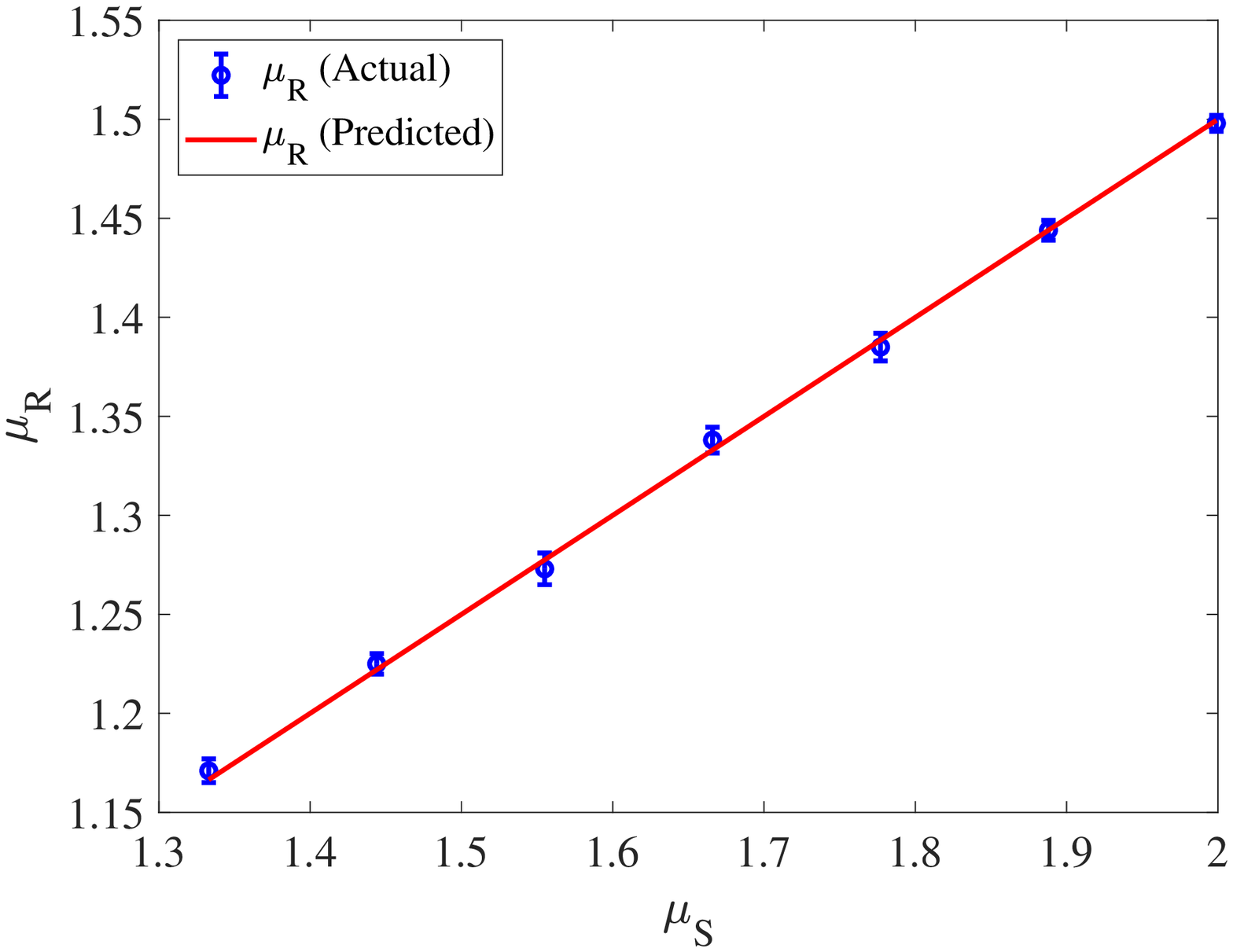}
\caption{$\mu_R$ evaluated numerically applying MDEA to the diffusion 
trajectories generated CTRW fluctuations with different values of $\mu_S$.} 
\label{fig2}
\end{figure}

In Fig. \ref{fig2} we establish this accuracy for different values of $\mu_S$. 
We generate a fluctuation $\xi(t)$ corresponding to a CTRW sequence with $\mu_S 
$ changing from $\mu_S = 1.3$ to $\mu_S = 2$. We turn these fluctuations into 
their corresponding diffusion trajectories. Then we apply MDEA to these 
diffusion trajectories to evaluate $\mu_R$. Fig. \ref{fig2} shows a very 
satisfactory agreement with the theoretical prediction of Eq. (\ref{casea}). 

\begin{figure}[ht]
\centering
\includegraphics[width=0.5\linewidth]{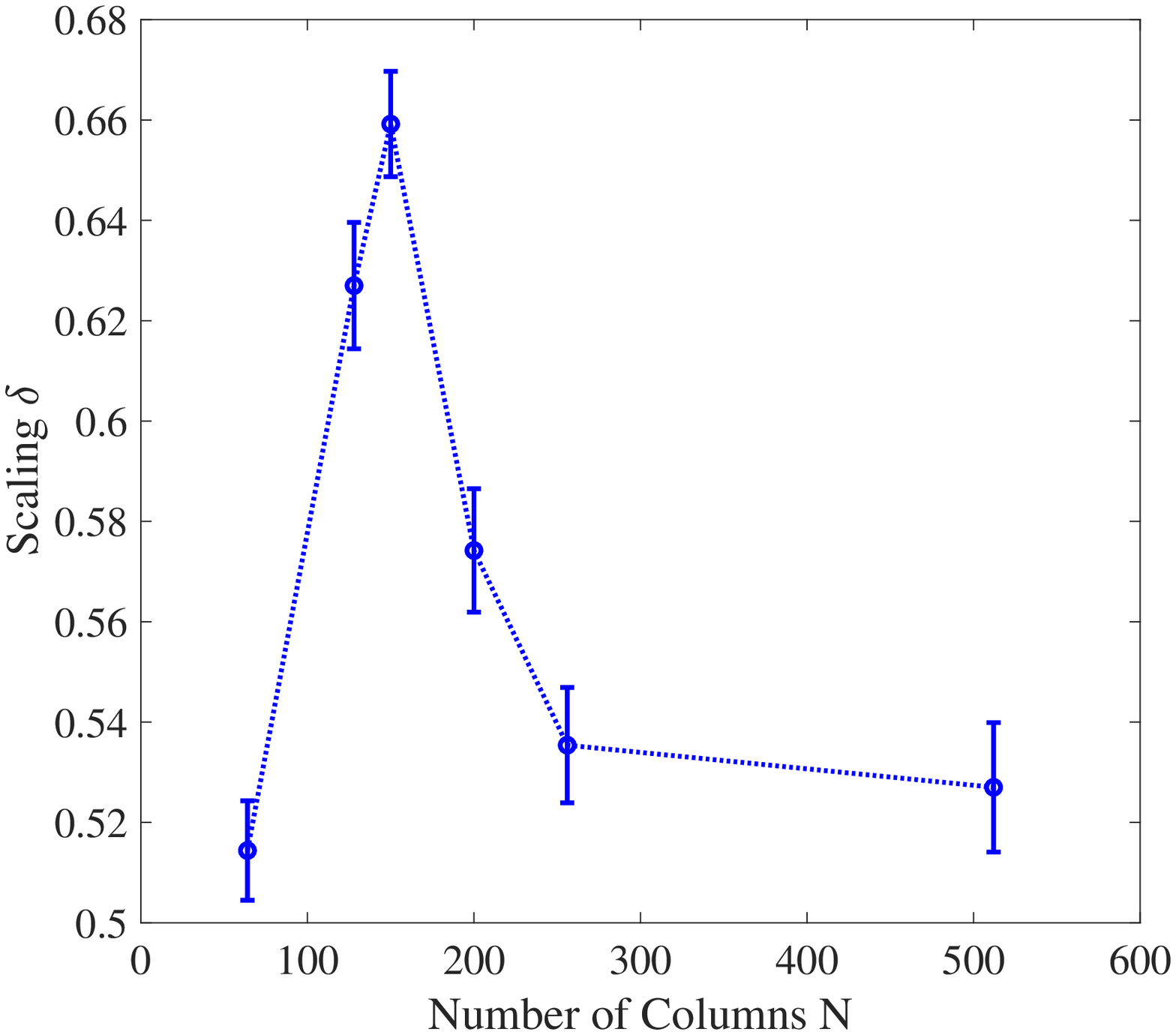}
\caption{Scaling of a column of the ballistic deposition model for a changing 
number $N$ of columns.} 
\label{fig3}
\end{figure}

\section{MDEA Applied to the Ballistic Deposition Model}

The Dunbar effect of this paper is based on Eq. (\ref{roberto}), which was 
originally proposed with no discussion of the Dunbar effect in Ref.
\cite{failla}.  To confirm the connection between the Dunbar effect and KPZ 
universality, we adopt the ballistic deposition model \cite{barabasi} and, as 
done in Ref. \cite{failla}, we observe the time evolution of single column, 
defining

\begin{equation} \label{letusmakemaurogreat}
\xi(t) = h(t) - \langle h(t) \rangle, 
\end{equation}
where $h(t)$ is the height of a randomly selected column. Due to the periodic 
boundary conditions adopted, different columns are equivalent. The value $
\langle h(t) \rangle$ denotes the mean height of the one-dimensional surface. 
$N$ is the number of columns. We apply to the analysis of $\xi(t) $ the MDEA, 
and we obtain the result of Fig. \ref{fig3}, which, as expected, yields a 
distinct Dunbar effect. 

\section{Derivation of $\bf \boldsymbol{\mu}_R = 1.75$}

\begin{figure}[ht]
\centering
\includegraphics[width=0.5\linewidth]{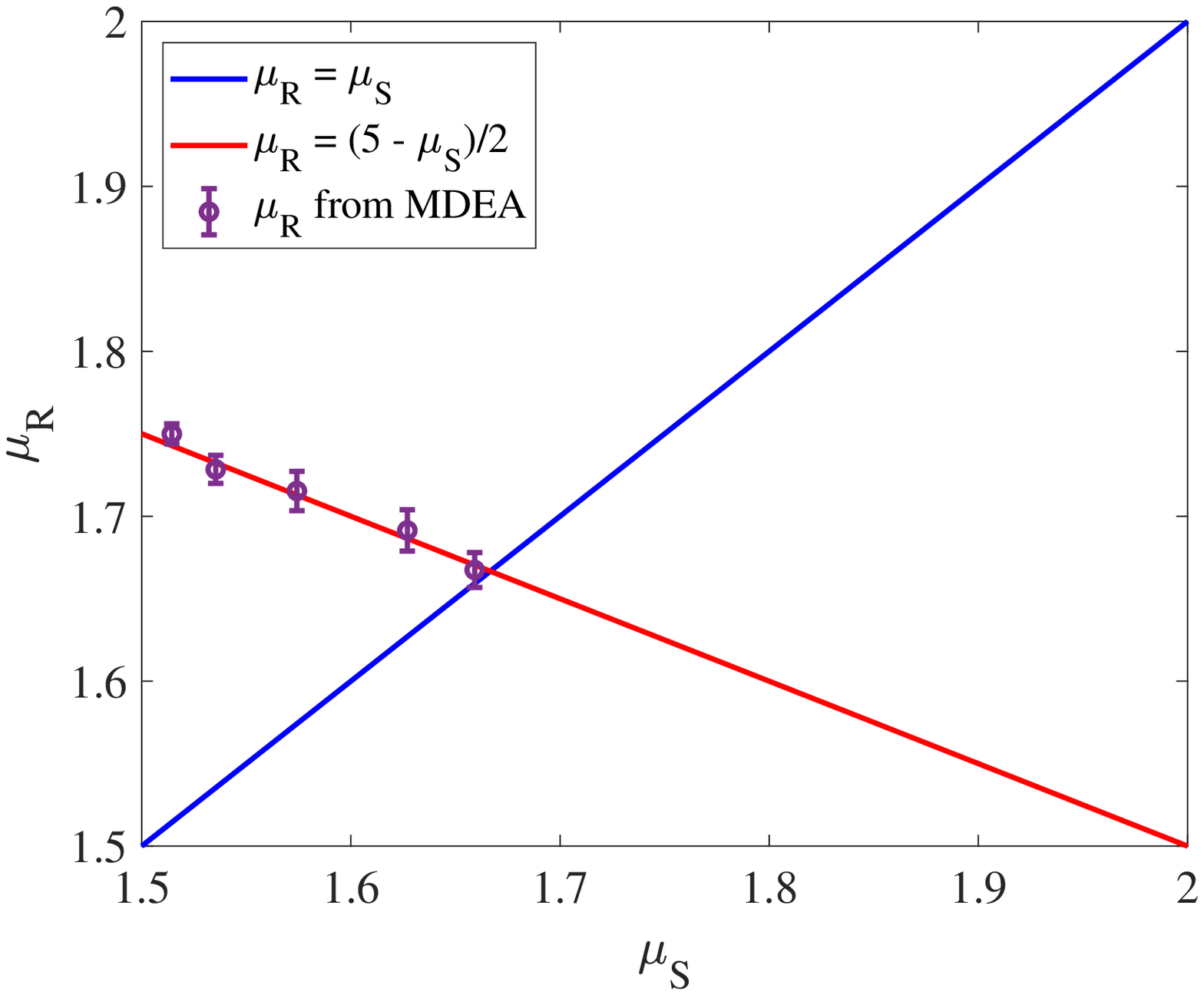}
\caption{Geometric derivation of the KPZ scaling. The red line illustrates the 
trajectory argument based on Eq. (\ref{roberto}). The blue line is the 
prediction of subordination to the fluctuation-dissipation process, $\mu_R = 
\mu_S$. The numerical dots are obtained by applying MDEA to the analysis of the 
diffusional trajectories $x(t)$ generated by the fluctuation 
$\xi(t)$ of Eq. (\ref{letusmakemaurogreat}) obtained from the  model of 
ballistic deposition \cite{failla}.} 
\label{fig4}
\end{figure}

Let us discuss the relation between $\mu_R$ and $\mu_S$, based on Eq. 
(\ref{roberto}).  In Fig. \ref{fig4} we plot Eq. (\ref{roberto}) red straight 
line) and $\mu_R = \mu_S$ (blue straight line). The red line at $\mu_S = 1.5$ 
holds the value $\mu_R = 1.75$ and it intersects the blue line at $\mu_S = 
1.67$.  Note that the numerical dots of this figure have been obtained by 
applying the MDEA to the diffusion trajectory $x(t)$ generated by the 
fluctuation $\xi(t)$ of Eq. (\ref{letusmakemaurogreat}). 

The left triangle of Fig. \ref{fig4} defines the operation region of Fig. 
\ref{fig3} of the text. We see that $\mu_R$ moves in the interval $[1.76, 1.67]
$ and that $\mu_R$ is always larger than $\mu_S$, thereby implying that the 
opinion persistence is less stable than the invisible crucial events.

\begin{figure}[ht]
\centering
\includegraphics[width=0.5\linewidth]{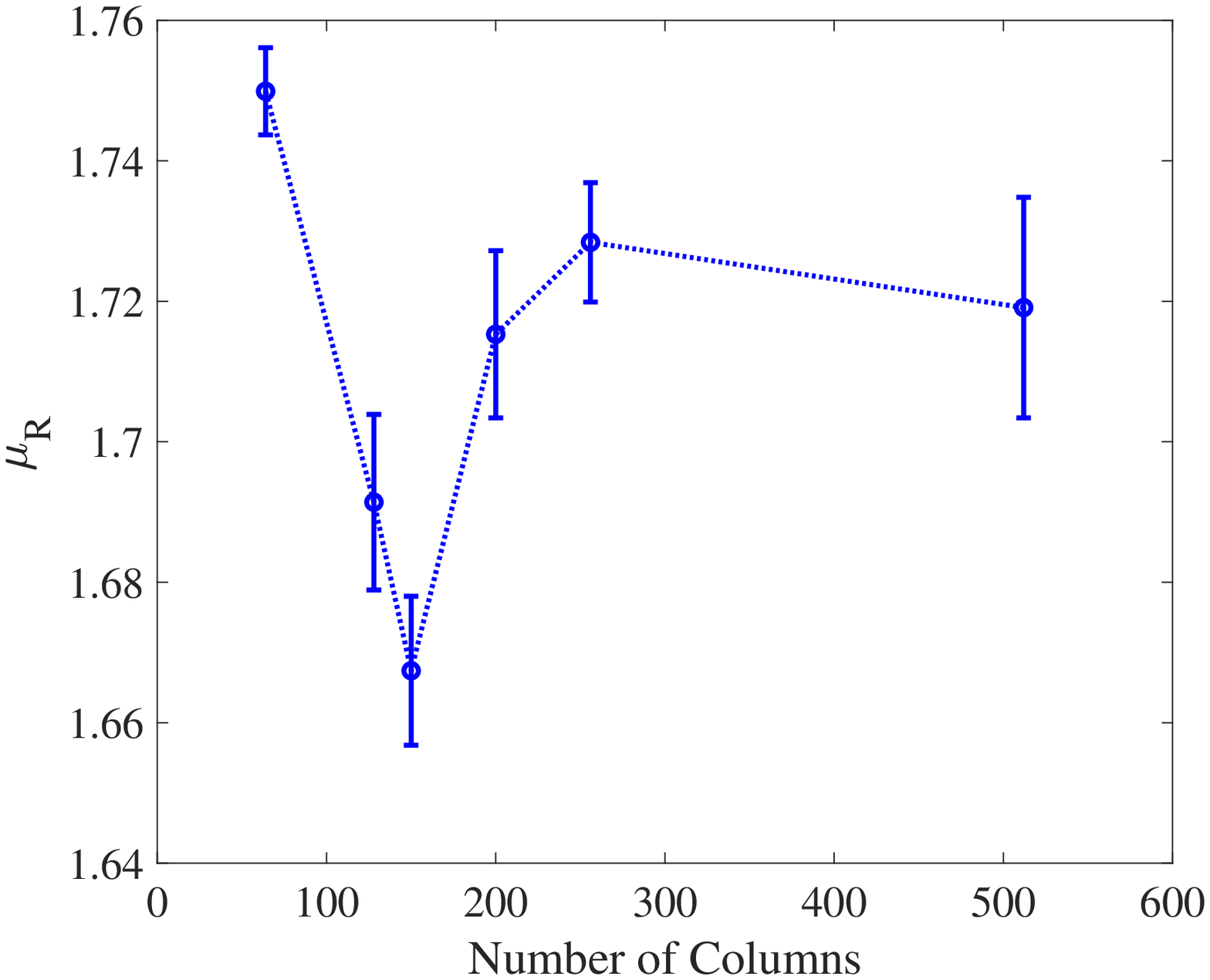}
\caption{$\mu_R$ as a function of N in the ballistic deposition model
\cite{barabasi}.}
\label{fig5}
\end{figure}

\section{MDEA applied to ballistic deposition data to evaluate $\bf \boldsymbol{\mu}_R$}

Fig. \ref{fig5} shows $\mu_R $ as a function of $N$ for the ballistic 
deposition model\cite{barabasi}. Again the drop of $\mu_R$ to the value $\mu_R 
= 1.67$ at $N =  150$ is remarkable. We stress that the intermediate 
asymptotics for N very large is very short, making the evaluation of the slope 
less accurate. This may be the cause of $\mu_R$ departing from the value $\mu_R 
= 1.75$. We cannot rule out the possibility that $\delta$ for $N$ large is  
larger than $0.5$, albeit significantly smaller than $0.67$. 

\section{Concluding Remarks}

The authors of Ref. \cite{failla} used the subordination to a fluctuation-dissipation process in the saturation regime of the ballistic deposition model. 
In this Letter we show that in this regime, 
\begin{equation}
\mu_R = \mu_S = 1 + 2 \delta
\end{equation}
This condition should be made compatible with the trajectory arguments, 
yielding.
\begin{equation}
\mu_R = 2-\delta.
\end{equation}
We see that these two equations are compatible when
\begin{equation}
\delta = \frac{1}{3},
\end{equation}
which is the KPZ scaling.

Note that the Dunbar point, $\mu_R = \mu_S = 5/3$, can also be reached, in 
principle, moving along the blue line of Fig. \ref{fig4}. The numerical results 
obtained using MDEA are accurate enough as to rule out this as a path to the 
Dunbar effect. The true path to the Dunbar effect is given by the red line of 
Fig. \ref{fig4}. As stressed in Ref. \cite{failla},  the dynamics of ballistic 
deposition are a self-organization  process spontaneously yielding the  
subordination to an ordinary fluctuation-dissipation process. Consequently,  as 
proved in this supplementary material, to $\mu_{S} = \mu_{R}$. This paper 
proves that this form of subordination is realized when $N = 150$. When $N \neq 
150$, Eq. ({\ref{roberto}) applies. 

\bibliography{supplementary}

%
%
%
